# Verification and Validation of a Numerical Wave Tank Using Waves2FOAM

**Chris Chartrand**[*]
Sandia National Laboratories
Albuquerque, NM, USA

**Nima Fathi**
University of New Mexico
Albuquerque, NM, USA

[*]Corresponding author: ccchart@sandia.gov

## 1. INTRODUCTION

Clean energy systems have been investigated recently by the researchers of the University of New Mexico to increase the efficiency specifically in the solar tower technology. Similar to solar energy, wave power harnesses energy that comes from the sun. Solar irradiation causes wind by changing the pressure, and wind gives its momentum to the ocean surface which produces waves [1-9]. Modeling and simulation (M&S) of free surface flow can be a very useful tool for the wave energy industry. However, if these tools are to be relied upon, their accuracy must be tested and shown to be sound. In this investigation, we focus on verification and validation (V&V) of wave tank flow M&S. The Volume of Fluid (VOF) model was applied to perform transient computational fluid dynamics (CFD) analysis using the numerical simulation software OpenFOAM. To assess V&V, the height of the free surface flow was considered as the system response quantity (SRQ). Stokes theory (fifth order) was utilized as the most accurate available approximation of the exact solution to measure the order of accuracy of the discretized mathematical model which is known as code verification [10, 11]. Data collected during model-scale testing at the Naval Surface Warfare Center, Carderock Division (NSWCCD) Maneuvering and Seakeeping (MASK) basin were utilized to validate the computational results [12]. The CFD domain consists of a two dimensional plane representing the region of interest of the wave basin which has several wave height measurement probes. Our simulation results were compared to the data from one of these probes in order to conduct the V&V analysis. Code verification and comparison with experimental values demonstrate the accuracy and efficiency of our model.

## 2. CARDEROCK WAVE TANK

The MASK wave basin was used to perform tests studying the motion of a floating buoy under the influence of surface waves [12, 13]. Although the experimental setup referenced here was focused on the motion of a floating body, baseline data was recorded of water surface elevation for an empty tank for tuning purposes. The V&V study here will use the empty tank data for comparison to numerical results.

The test equipment consists of a 60 m × 100 m wave basin filled to a water depth of 4 m. A pivoting bridge spanned the wave basin length, and measurement devices were mounted to the bridge. In the current experiments, the bridge was pivoted at approximately a 70° angle and aligned with the direction of wave propagation. A two-dimensional sinusoidal wave with an amplitude of 0.05 m, a and a period of 2.5 s was generated at the paddles of the device, and propagating in the direction parallel to the bridge.

## 3. NUMERICAL SIMULATION

### 3.1 Simulation Setup

Using the Waves2FOAM package [14], a two dimensional numerical wave was simulated corresponding to the experimental wave conditions. Waves2FOAM is an OpenFOAM package which utilizes the Volume of Fluids approach to model the free-surface of wave flows. A sinusoidal inflow wave condition was set to exactly match those of the experiment with the amplitude of 0.05 m and a period of 2.5 s. The numerical domain covered an 8 m x 105 m (H x L) region with the initial water level sitting at the height of 4 m. The 105 m length of the domain consisted of a 5 m wave generation region, followed by an 85 m wave propagation region, and then a 15 m wave absorption region. The total simulation time was 250 s. Given the wavelength of 9.662 m and the period of 2.5 s, the calculated wave speed for this scenario is 3.86 m/s. This would imply that the wave generated at the inflow of the tank should propagate through the 85 m domain in approximately 22 s. To ensure that all artifacts of the initialization wave had washed through, only simulation data after 100 s of simulation time was used for analysis. In order to test the error convergence, a set of simulations was set up using different levels of discretization in both space and time, all with identical flow conditions. The three spatial discretization used in the runs were $\Delta x = 1.0, 0.5, 0.2$. The temporal discretization used were $\Delta t$ = 0.02, 0.1, 0.05 s.

## 3.2 Numerical Wave Results

Here are the results of the simulation setup described in Subsection 3.1. A probe approximately 75 m from the wave generation region was used for comparison. The SRQ in this study was water surface elevation. The OpenFOAM solution uses second order discretization schemes indicating an expected order of accuracy of 2 for the velocity of converged steady state flows.

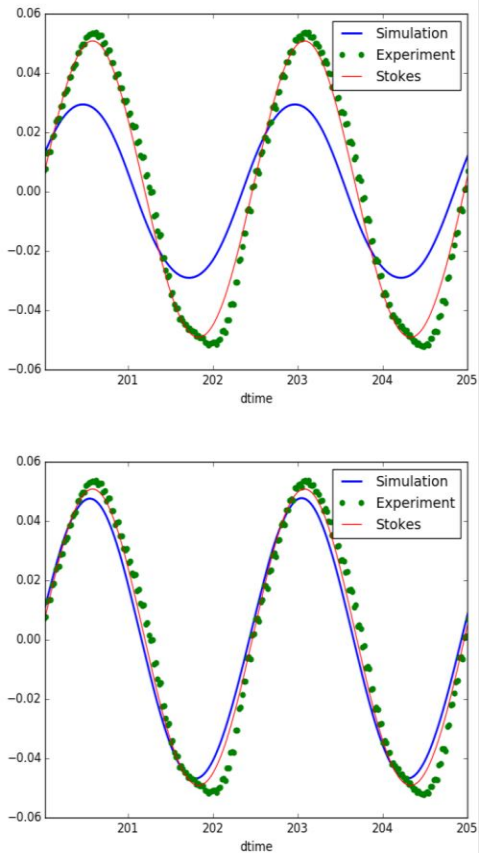

**FIGURE 1. COMPARISON OF SIMULATION TO EXPERIMENTAL RESULTS FOR $\Delta x = 1.0, 0.25m$ (TOP TO BOTTOM) AND $\Delta t = 0.02s$.**

However, surface elevation is not a variable directly solved for in the system of equations used by OpenFOAM. Rather it is a post processed quantity obtained from the water percentage solved for in the free surface interface cells. The resulting order of accuracy in the surface elevation may not be straightforward.

Figures 1 and 2 show the water surface elevation solution from diffidently discretized solutions as compared to the experiments and analytical solution. The plots in Figure 1 clearly depict an increase in accuracy with finer spatial discretization while the change in autocracy is not as apparent in the temporal discretization comparisons of Figure 2. A similar trend is seen in the plots of the FFTs in Figures 3 and 4. Below we look to quantify the error and determine the spatial and temporal order of accuracy.

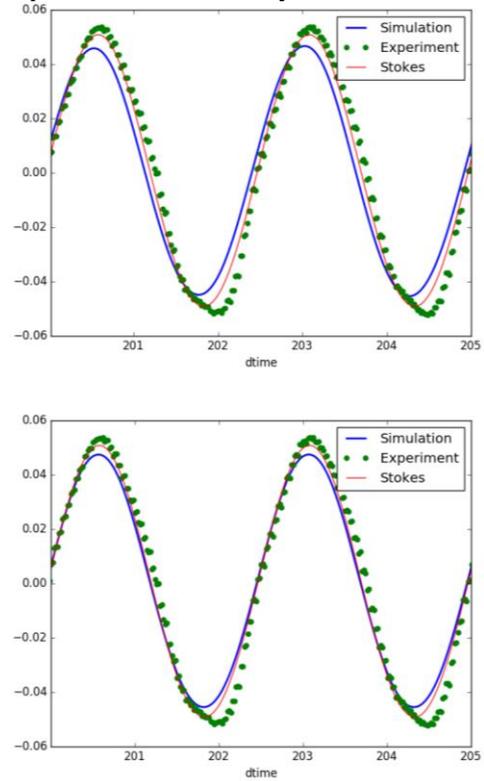

**FIGURE 2. COMPARISON OF SIMULATION TO EXPRIMENTAL RESULTS FOR $\Delta x = 0.5m$ AND $\Delta t = 0.02, 0.05s$ (TOP TO BOTTOM).**

### 3.3 Error Calculation

The transient nature of this problem introduces complications in determining the solution error. In this work, we define the solution error as follows.

$$\varepsilon_{h,g} = \int_{a}^{a+\tau} |\bar{u}(t) - u_{h,g}(t)| dt \quad (1)$$

Here, $\bar{u}(t)$ represents the exact solution as a function of time. The value of $u_{h,g}(t)$ represents the numerical solution as a function of time with a given spatial ($h$) and temporal ($g$) discretization. The integral represents the total difference between the numerical solution and the exact solution over a full wave period ($\tau$). This is the value we will use to represent our solution error.

Table 1 below shows the numerical error for the nine discretization combinations run in the simulation matrix. Figure 5 represents these values on a log-log plot to show the order of convergence. The plots show that for the finest temporal discretization, the spatial order of accuracy of the surface evaluation approaches approximately 1.0, while for the finest spatial discretization, the temporal order of accuracy approaches approximately 0.75.

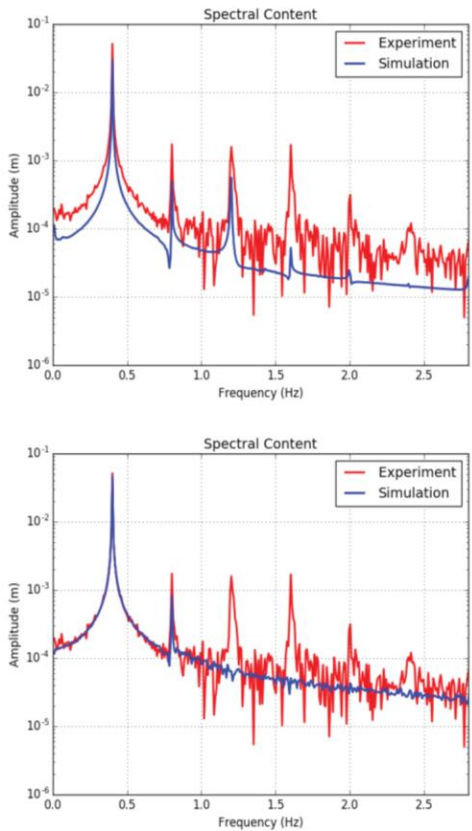

FIGURE 3. COMPARISON OF SIMULATION TO EXPERIMENTAL FFTS FOR $\Delta x = 1.0, 0.25\,m$ (TOP TO BOTTOM) AND $\Delta t = 0.02s$.

## 4. CONCLUSION

The numerical wave tank model demonstrates here has been verified for wave surface height modeling against a fifth-order Stokes theory, showing a very good agreement between the simulation and the analytic solution for a propagating sinusoidal wave. The model has also been validated as effective at accurately representing the physical phenomena observed in an experimental facility, with the plots of wave surface elevation showing a good qualitative agreement between the simulation results and experiments as well. In addition, the FFTs of simulation results compare quite well with those of the experimental results, capturing the first two, and sometimes third, tones of the wave spectrum.

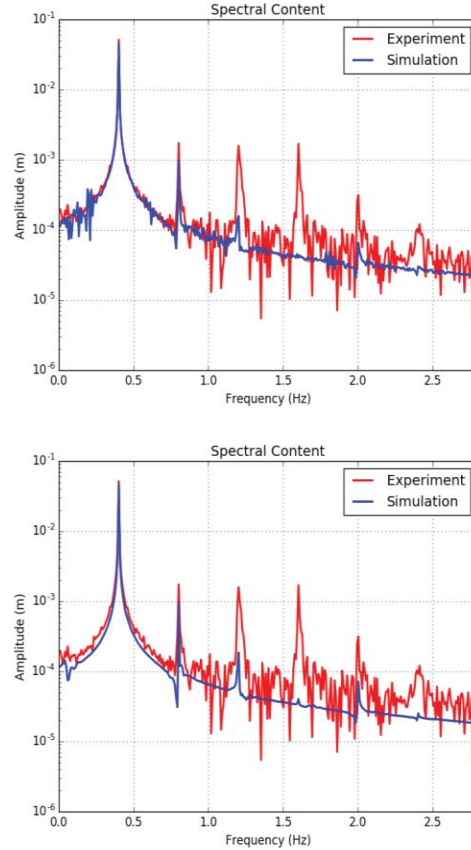

FIGURE 4. COMPARISON OF SIMULATION TO EXPERIMENTAL FFTS FOR $\Delta x = 0.5\,m$ AND $\Delta t = 0.02, 0.05\,s$ (TOP TO BOTTOM).

The observed order of accuracy was not shown to be as high as expected for the model. The OpenFOAM simulations were performed using a second order spatial scheme and a first-order temporal scheme. This means that the observed order of accuracy is lower than the expected in both space and time.

There may be valid reasons for this result, however, firstly, while the analytic Stokes solution is a very high order approximation, it's not strictly the exact solution Secondly, the temporal nature of this problem makes standard error calculations difficult. Slight shifts in wave phase can have a large impact on the error calculation, due to the method used to subtract the two solutions.

**TABLE 1. : TABLE OF SIMULATION ERROR (COMPARED TO STOKES) AS A FUNCTION OF SPATIAL AND TEMPORAL DISCRETIZATION**

| $\Delta x$ | $\Delta t$ | Error |
|---|---|---|
| 1.0 | 0.02 | 1.4762e-02 |
| 0.5 | 0.02 | 5.2156e-03 |
| 0.25 | 0.02 | 3.3988e-03 |
| 1.0 | 0.01 | 1.2625e-02 |
| 0.5 | 0.01 | 3.9637e-03 |
| 0.25 | 0.01 | 1.8332e-03 |
| 1.0 | 0.005 | 1.2424e-02 |
| 0.5 | 0.005 | 2.1972e-03 |
| 0.25 | 0.005 | 1.0899e-03 |

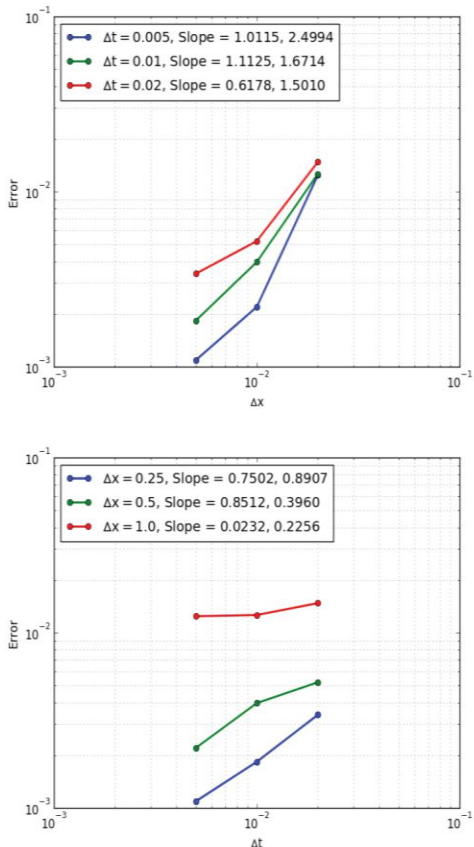

**FIGURE 5. NUMERICAL ERROR VS $\Delta x$ (TOP) AND $\Delta t$ (BOTTOM) FOR EACH TEMPORAL DISCRETIZATION.**

**ACKNOWLEDGEMENTS**

This work was funded by the U.S. Department of Energy's Water Power Technologies Office. Sandia National Laboratories is a multi-mission laboratory managed and operated by Sandia Corporation, a wholly owned subsidiary of Lockheed Martin Corporation, for the U.S. Department of Energy's National Nuclear Security Administration under contract DE-AC04-94AL85000.